\documentclass{nature}
\usepackage{graphicx}
\usepackage{xcolor}
\usepackage{amsmath}
\usepackage{amssymb}
\usepackage{rotating}
\usepackage{url}
\usepackage{setspace}
\usepackage{float}
\usepackage{threeparttable}
\usepackage{xcolor}

\makeatletter
\let\saved@includegraphics\includegraphics
\AtBeginDocument{\let\includegraphics\saved@includegraphics}
\renewenvironment*{figure}{\@float{figure}}{\end@float}
\makeatother

\def \msol {M$_{\odot}$ }
\def \msols {M$_{\odot}$}
\def \kmss {km s$^{-1}$}
\def \kms {km s$^{-1}$ }

\begin{document}

\title{H$_2$O MegaMaser emission in NGC 4258 indicative of a periodic disc instability}

\author{Willem A. Baan$^{1,2,3}$, 
Tao An$^3$, 
Christian Henkel$^{4,5,1}$, 
Hiroshi Imai$^6$, 
Vladimir Kostenko$^7$, 
and Andrej Sobolev$^{8}$ 
}
\maketitle

\begin{affiliations}
\item Xinjiang Astronomical Observatory, Chinese Academy of Sciences, 150 Science 1-Street, Urumqi, Xinjiang 830011, China
\item Netherlands Institute for Radio Astronomy ASTRON, Oude Hoogeveensedijk 4,  7991 PD Dwingeloo, The Netherlands, email: baan@astron.nl
\item Shanghai Astronomical Observatory, Chinese Academy of Science, Nandan Road 80, Shanghai 200030, China

\item Max-Planck-Institut f\"ur Radioastronomie, Auf dem H\"ugel 69, D-53121 Bonn, Germany
\item Astron. Dept., King Abdulaziz University, P.O. Box 80203, Jeddah, Saudi Arabia
\item Amanogawa Galaxy Astronomy Research Center, Graduate School of Science and Engineering,  Kagoshima University, 1-21-35 Korimoto, Kagoshima 890-0065, Japan
\item AstroSpace Centre, Lebedev Institute, Moscow, GSP-7, 117997, Russia
\item Astronomical Observatory, Ural Federal University, Lenin Ave. 51, Ekaterinburg 620083, Russia  
\end{affiliations}

{\it Published in Nature Astronomy, Volume 6, p. 976-983, June 2022}
\vspace{10mm}

\begin{abstract}
H$_2$O MegaMaser emission may arise from thin gas discs surrounding the massive nuclei of galaxies such as NGC\,4258, but the physical conditions responsible for the amplified emission are unclear. 
A detailed view of these regions is possible using the very high angular resolution afforded by space very long baseline interferometry (SVLBI). 
Here we report SVLBI experiments conducted using the orbiting RadioAstron Observatory that have resulted in detections of the H$_2$O 22 GHz emission in NGC\,4258, with Earth-space baselines of 1.3, 9.5 and 19.5 Earth diameters.
Observations at the highest angular resolution of 11 and 23 $\mu$as show distinct and regularly spaced regions within the rotating disc, at an orbital radius of about 0.126 pc.
These observations at three subsequent epochs also indicate a time evolution of the emission features, with a sudden rise in amplitude followed by a slow decay.
The formation of the emission regions, their regular spacing and their time-dependent behaviour appear consistent with the occurrence  of a periodic magneto-rotational instability in the disc.
This type of shear-driven instability within the differentially rotating disc has been suggested to be the mechanism governing the radial momentum transfer and viscosity within a mass-accreting disc. 
The connection  of the H$_2$O MegaMaser activity with the magneto-rotational instability activity would make it an indicator of the mass-accretion rate in the nuclear disc of the host galaxy. 
\end{abstract}

{\bf 1. Introduction}

The Space Radio Telescope (SRT) of the RadioAstron Observatory (RAO) is an observing station onboard the Spectr-R 
spacecraft on a highly elliptical orbit with an apogee of 350 000 km\cite{KardashevEA2013}. 
Operating this SRT in combination with ground-based radio telescopes allows for Space Very Long Baseline Interferometry 
(SVLBI) observations with baselines up to thirty times as long as those possible with Earth-based telescope arrays and a 
spatial resolution at the target sources 30 times higher. 
Such high-resolution observations enable the study of the structural details of the H$_2$O maser emission in the nearby 
(7.6 Megaparsec) galaxy NGC\,4258 (also named Messier 106). 

Many extragalactic H$_2$O MegaMaser (MM) hosting galaxies show line emission associated with extended and shocked 
regions in their interstellar medium, possibly as a result of mergers or interactions of jet outflows with the ambient medium.
However, the H$_2$O emission in the MM NGC\,4258 originates within the fast rotating disc surrounding the active galactic 
nucleus (AGN) and it has become a prototype for about one third of those 
among the 160 known MM sources\cite{Lo2005,ZhuEA2011,Braatz2014,KuoEA2020}.
The behaviour of the maser emission regions in NGC\,4258 has been extensively studied over many years\cite{HaschickB1994,MiyoshiEA1995,GreenhillEA1995,HerrnsteinEA1999,HerrnsteinEA2005}.
The maser regions have been well modelled with an edge-on and warped thin Keplerian disc resulting
in the most accurate estimate of the mass of the central black hole of 4.00 $\pm$0.09 $\times$ 10$^7$ \msols, and a very 
accurate distance to the galaxy of 7.6 $\pm$0.17 $\pm$0.15 Mpc, with a systemic line-of-sight velocity of 474.3$\pm$0.5 \kms 
in the local standard of rest (LSR)\cite{HerrnsteinEA1999,ArgonEA2007,HumphreysEA2008,HumphreysEA2013}.
The H$_2$O maser emission originates at three locations in the disc of NGC\,4258, at the systemic velocity of the galaxy, 
with orbiting molecular regions passing from west to east in front of the nuclear region and at the blue- and redshifted 
edges of the disc.
The emission at the systemic velocity is most prominent because there the warped disc lies along 
the line of sight with the north-south jet extension from the AGN in NGC\,4258\cite{HerrnsteinEA1998,ArgonEA2007,DoiEA2013}, 
which facilitates maser amplification of a background radio continuum by excited H$_2$O molecules in the foreground section 
of the disc\cite{Baan1985}.
Maser features close to the systemic velocity continuously drift upwards in velocity as they transit from 
the approaching side to the receding side of the disc; i.e. from below to above the systemic velocity.
The drift rate at which this happens depends on the orbital velocity and the distance from the central nucleus. 
The emission in the approaching western and the receding eastern edges of the disc is much weaker, likely because 
of the absence of substantial background continuum. 
However, these edge regions also define the approximate radial extent within the disc from 0.14 parsec (pc) to 0.29 pc 
where the masering activity can occur\cite{HerrnsteinEA2005,HumphreysEA2013}.

{\bf 2. Results and Discussion}

\begin{figure}[h]
\includegraphics[scale=0.8,angle=0]{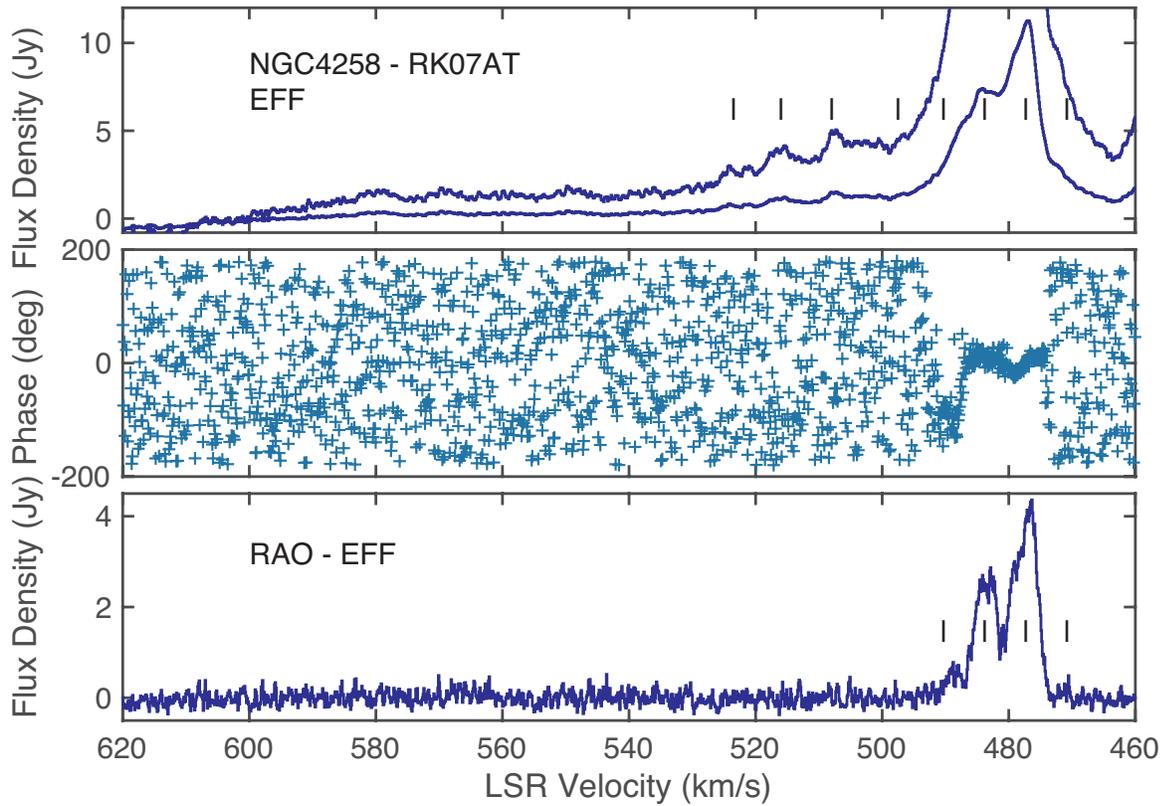}
\caption{{\bf The first experiment RAKS07AT for NGC\,4258 with a 9.1 - 9.8 ED Earth-space baseline taken on 05.02.2014.}
(Top) The Stokes $I$ auto-correlation flux density spectrum from the Effelsberg Telescope (EFF).
A bandpass calibrator was used for removing the baseline. 
An amplified (factor 8) second trace shows the presence of additional weak periodic components up to 530 \kmss.
(Middle) The interferometric fringe visibility phases of the RAO-EFF Stokes $I$ cross-correlation spectrum.
(Bottom) The RAO-EFF Stokes $I$ cross-correlation flux density spectrum showing four spectral components.  
A phase jump may be seen in the middle of both features.
\label{fig1}}
\end{figure}

\begin{figure}[h]
\includegraphics[scale=0.8,angle=0]{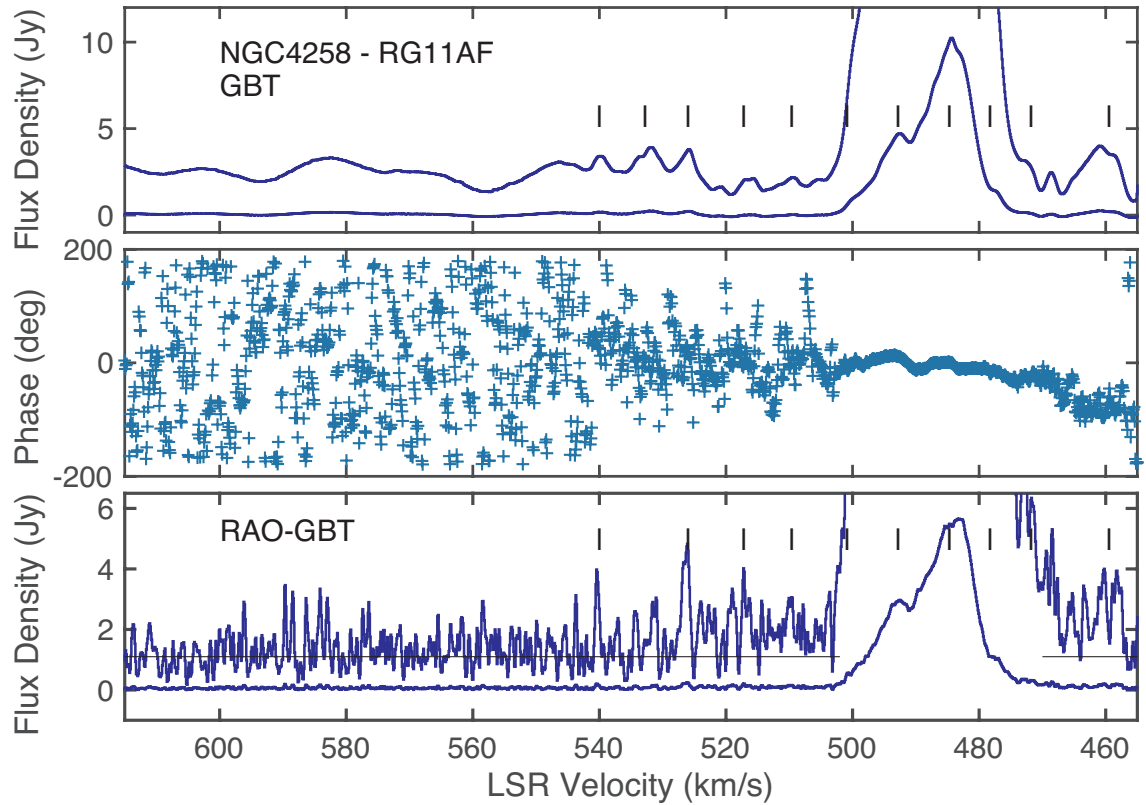}
\caption{{\bf Experiment RAGS11AF for NGC\,4258 with a 0.5 -1.9 ED Earth-space baselines taken on 18.12.2014.}
(Top) The Stokes $I$ auto-correlation flux density spectrum from the Green Bank Telescope (GBT).
A second amplified (factor 8) trace shows the higher velocity auto-correlated components extending to 540 \kmss.
No bandpass calibrator was available for this observation and a fitted baseline was removed from the spectrum.
(Middle) The interferometric fringe visibility phases of the RAO-GBT Stokes $I$ cross-correlation flux density spectrum.
(Bottom) The RAO-GBT Stokes $I$ cross-correlation flux density spectrum.
A second trace presents an amplified (factor 20) profile showing the presence of weaker cross-correlated features.
\label{fig2}}
\end{figure}

\begin{figure}[h]
\includegraphics[scale=0.8,angle=0]{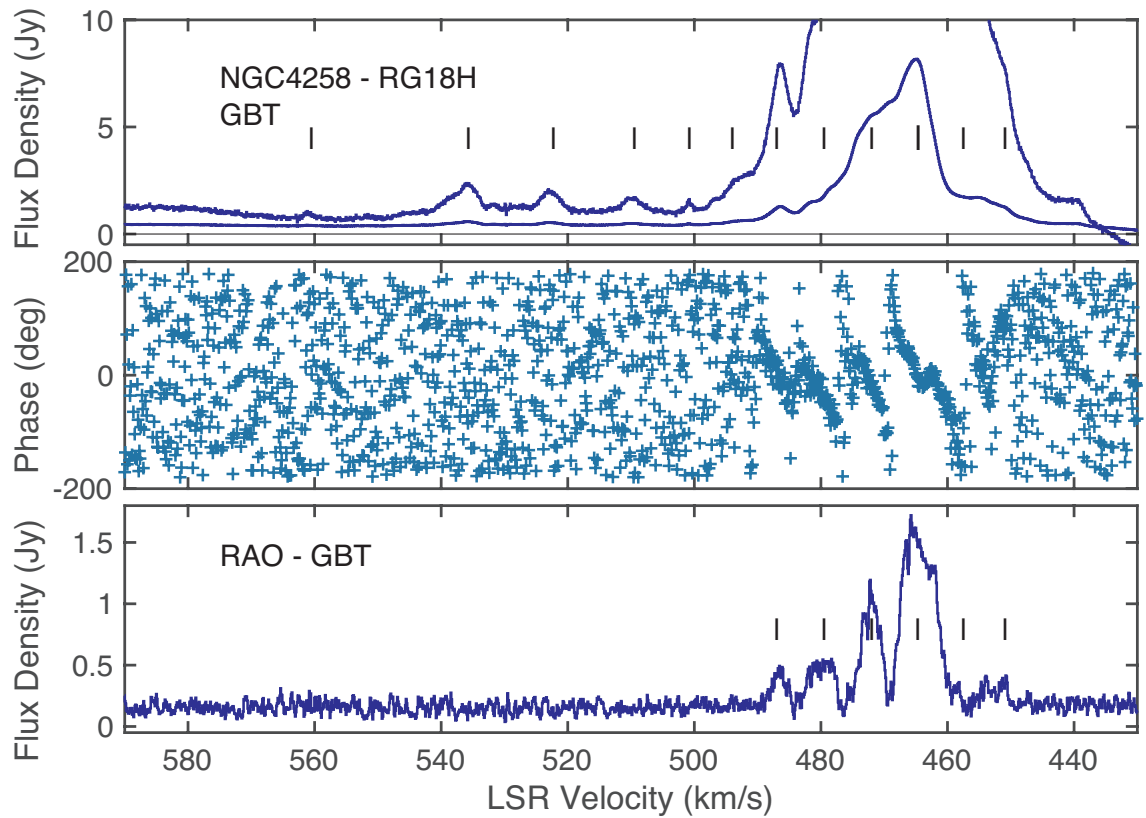}
\caption{{\bf Experiment RAGS18H for NGC\,4258 with a 19.5 ED Earth-space baseline taken on 17.03.2016.}
(Top) The Stokes $I$ auto-correlation flux density spectrum for the Green Bank Telescope.
An amplified (factor 8) second trace of the profile shows the presence of weak high velocity features.
A bandpass calibrator was used for removing the spectral baseline. 
(Middle)  The interferometric fringe visibility phases  of the RAO-GBT Stokes $I$ cross-correlation spectrum.
(Bottom) The RAO-GBT Stokes $I$ cross-correlation flux density spectrum showing at least five well-spaced features. 
A phase jump may be seen in the middle of the strongest feature at 464 \kms and the feature at 480 \kmss. 
\label{fig3}}
\end{figure}

The results of three SVLBI observations of the systemic 22 GHz H$_2$O emission of NGC\,4258 
are presented in the form of total power (single dish) and cross-correlated spectra. 
Imaging of the emission regions is not (yet) possible with a limited number of antennae and such extremely long baselines. 
The Key Science Project experiment RAKS07AT on Earth-space baselines of 9.1 -- 9.8 Earth Diamaters (ED)  was conducted 
on 5 February 2014 (2014.099), giving a resolution or beam size of 23 $\mu$as or 175 Astronomical Units (AU) at the target (Figure \ref{fig1}). 
The General Observing Time GOT) experiment RAGS11AF with Earth-space baselines of 1.4 -- 1.9 ED was conducted on 18 December 2014 (2014.964),
giving a beam size of 113 $\mu$as or 861 AU) at the target  (Figure \ref{fig2}).
The third experiment, a GOT experiment RAGS18H on an Earth-space baseline of 19.5 ED  was conducted on 17 March 2016 (2016.210), giving a 
beam size of 11 $\mu$as or 84 AU at the target (Figure \ref{fig3}).
All experiments have been correlated with the ASC Software Correlator\cite{LikhachevEA2017}.

Although these experiments were done roughly one year apart, the auto-correlation spectra at these epochs show a 
similar broad-structured H$_2$O emission line together with a number of weaker components at higher velocities. 
The cross-correlated spectra and interferometric fringe-visibility phases from the three experiments show that at higher spatial resolution the 
broad emission profile breaks up into well-spaced features.
Two dominant components are detected on the shortest Earth-space baselines (Fig. \ref{fig2}),  
four well-spaced components are seen on an intermediate baseline (9.3 ED) (Fig. \ref{fig1}), 
and a complex profile with six well-spaced components is revealed on the longest baseline (19.5 ED) (Fig. \ref{fig3}).
At high resolution the spectrum shows well-spaced emission components that imply the existence of a string of 
compact and high brightness regions in the accretion disc. 
At lower resolution and with single dish experiments, these features are less visible because the spectrum is 
convolved with diffuse and halo emissions within the broader beam.
The flux density of these cross-correlated features decreases from about 60\% to 18\% of the auto-correlated 
flux density when going from 1.3 to 19.5 ED baselines.
 
\begin{figure}[h]
\includegraphics[scale=0.50,angle=-90]{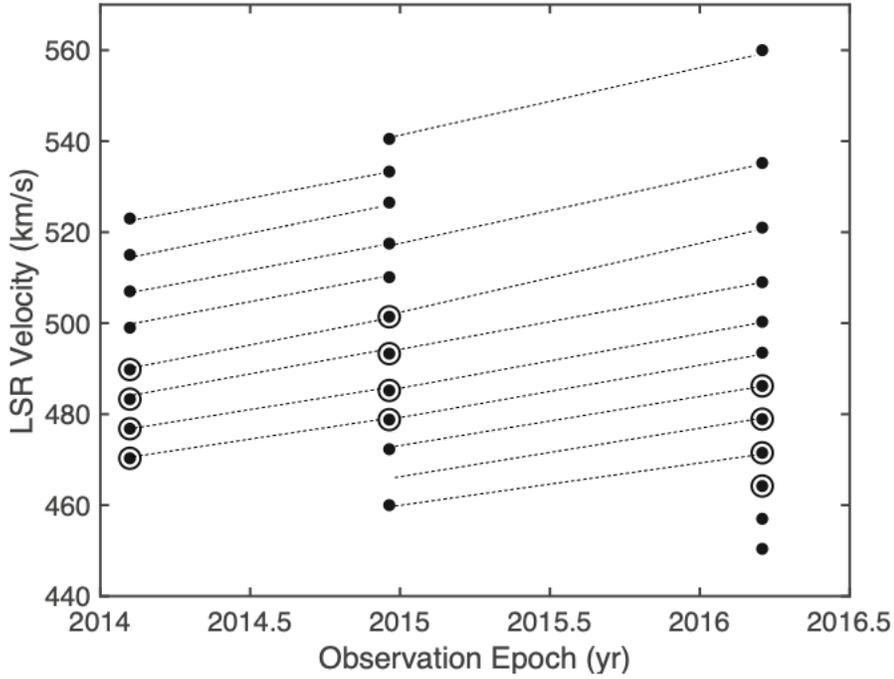}
\caption{{\bf Component velocities and velocity drifts between three epochs.}
All identified velocity components at the three epochs are displayed and the high brightness components 
have been indicated with a larger circle symbol.
Linear connections between the data points from RAKS07AT, RAGS11AF, and RAGS18H indicate the drift pattern of weak and strong velocity components as features re-appear at the next epoch.
A mean value of the velocity drift rate based on the linear extrapolations between the first and last experiments is 11.1 \kms yr$^{-1}$.
The three epochs also show the intensity evolution of individual velocity features with time, where the strong features of 
RAKS07AT and RAGS11AF show a decay and the lowest velocity features in the RAGS11AF spectrum evolve into prominent 
features in the RAGS18H spectrum (compare Figs. \ref{fig1} - \ref{fig3}).
Remnant features remain in the spectrum for years. 
\label{fig4}}
\end{figure}

The LSR velocity range of 430 - 560 \kms covered by the emission features is the same as that found in earlier monitoring 
experiments\cite{HaschickB1994,HerrnsteinEA2005,ArgonEA2007}.
However, the pattern of a multi-component broad feature trailing behind a series of weak features differs from the data 
obtained at earlier epochs where prominent groups of features and individual features were observed filling the whole 
velocity range. 
Analysis of the identifiable features in the current auto- and cross-correlated spectra shows that there is a nearly 
regular spacing in velocity of the prominent cross-correlated features of RAKS07AT and RAGS18H (see Fig. \ref{fig4} 
and Table \ref{tab1} in Methods). 
In particular, a mean velocity separation of 7.3$\pm$0.2 \kms is found for the RAGS18H data, while the RAKS07AT and 
also RAGS11AF data at earlier epochs show a separation of about 6.5$\pm$0.2 \kmss. 
The regular spacing of the main features also extends among the weaker features at higher velocities, although with some 
irregularity in the spacing and some undetected features.
This observed spacing and even those irregularities suggest that some ongoing dynamical process leads to the formation of the 
emission regions with an apparent periodicity. 

Another important property that further qualifies the nature of the masering regions is the velocity drift of the emission features over time 
that has been observed during previous monitoring observations \cite{HaschickB1994,HumphreysEA2008}.
The current observations also show that the velocity pattern of weak and strong features observed at one epoch 
repeats at a higher velocity during the next epoch, so that even the weak feature at the lowest velocity in the RAGS11AF spectrum 
corresponds to a prominent feature in the RAGS18H spectrum (Figs. \ref{fig1}-\ref{fig3}).
An analysis of the feature velocities in the current (systemic velocity) spectra at three separated epochs confirms a steady 
velocity drift of $\ddot R_d$ = 11.1$\pm$0.5 \kms yr$^{-1}$ (Fig. \ref{fig4}; Methods).
The observed drift velocity of the features is important because it confirms that the masering regions are in orbital motion 
and it also determines the radial location of those regions within a Keplerian disc.
The orbital velocity $V_k$ of an emission region moving within the Keplerian disc at a radius $R_{em}$ may be expressed as 
$V_k = 416\,R_{em}^{-1/2} M_4^{1/2}$ \kmss, while its associated velocity drift rate, or the orbital acceleration, may be expressed 
as  $\ddot R_d = V_k^2  R^{-1}_{em} = 0.176  M_4  R^{-2}_{em}$ \kms yr$^{-1}$. 
The observed velocity drift rate of systemic emission regions in the disc surrounding a central AGN mass $M_4 = 4 \times 10^7$ \msol 
identifies an orbital velocity of the regions of 1172 \kms and an orbital radius of $R_{em}$ = 0.126 pc. 
The current value for the velocity drift rate falls within the range 6.2 to 11.6 \kms yr$^{-1}$ found during earlier monitoring, which 
corresponds to a range in radius of 0.168 to 0.123 pc within in the disc\cite{HaschickB1994,HumphreysEA2008}. 
Apparently, the observed emission regions are close to the inner edge of the known molecular zone in the disc.

On this orbit the mean velocity separation for the prominent emission regions would indicate a geometric separation 
on the order of 172 AU, which may be compared with twice the scale height in the disc. 
For NGC\,4258 the local scale height in a standard thin accretion disc\cite{FrankKR1992} $H = c_s R_{em}^{3/2} (GM_{bh})^{-1/2}$, where $c_s$ is the local speed of sound,  
will be 63 AU at radius $R_{em}$ = 0.126 pc for a gas temperature of 1000 K, which agrees with the estimate of 
$H \leq 0.002 R_{em}$ = 62 AU suggested by ground-based observations\cite{HerrnsteinEA2005}.
The full width at half power of the best resolved features in the RAKS07AT and RAGS18H spectra indicates a size scale for the emission regions on the order of 95 AU, 
which is similar to the scale height of the disc and nearly half the separation between the regions.
The half power beam width is also similar to the scale size of 67 AU of the (related) absorbing structures found in the X-ray data of NGC\,4258\cite{FruscioneEA2005}. 
For comparison, the estimated half-thickness of the disc inferred from ground-based facilities of 190 AU is about three times the 
estimated scale height of the disc\cite{ArgonEA2007}.

The current understanding of the H$_2$O emission in NGC\,4258  is that the emission regions occur in a radial zone within 
the rotating disc where viscous heating and external X-ray illumination can forge a transition from atomic to molecular gas\cite{NeufeldMC1994,NeufeldM1995,MaozM1998,HerrnsteinEA2005,Lo2005}.
When the density in this radial zone is in the required range of 10$^8$ to 10$^{10}$ cm$^{-3}$ and the temperature in the range of 
400 - 1\,000 K\,, collisional excitation of the water molecules can result in a population inversion and a negative optical depth. 
Regions with a negative optical depth of the H$_2$O population in the front section and in any edge-sections of the masering disc may 
then result in maser-amplification of background radio continuum and observable maser features at the 
systemic velocity and at high- and low-velocities\cite{Baan1985}.
The systemic emission features in NGC\,4258 have been associated  with cloud-cloud superpositions or multi-armed spiral 
patterns in the disc and/or the presence of turbulent cells within this molecular 
ring\cite{WallinWW1998,MaozM1998,WallinWW1999,KartjeKE1999,BabkovskayaV2000}. 
Similarly the well-spaced (and possibly periodic) high-velocity features may result from the alignment of spiral shock 
waves\cite{MaozM1998}, turbulent cells in the molecular ring \cite{KartjeKE1999}, or flexural waves in the warping 
disc\cite{BabkovskayaV2000}.

The observational data for NGC\,4258 from these experiments indicate that the strong masering features first appear 
at the (apparent) low-velocity end of a string of spectral features drifting upward in velocity over time (moving leftwards in 
Figs. \ref{fig1}-\ref{fig3}) and that features re-appear at a higher velocity in subsequent observing epochs.
Because the spectra are actually snapshots of the ongoing masering activity, the sequence of spectra at three epochs show a time 
evolution of the features. 
After appearing at lower velocities, the features show a very rapid initial rise in strength followed by a slower decay
on time scales of years, after which  weak and decaying features remain for a longer time as they drift towards higher velocities.
Assuming that maser amplification is responsible for the emission from within the disc, a velocity-aligned and line-of-sight column 
density of excited molecules will be superposed on the background radio continuum.
Assuming that this background continuum at the base of the southern jet is rather uniformly distributed (Methods),
this velocity-aligned column density needs to be specially generated because the differential rotation in the disc prevents any such alignment.
Some mechanism is required to structurally perturb the disc and produce a strong temporary enhancement of the amplifying optical depth
to facilitate longer-lasting and apparently regularly-spaced emission regions.
The apparent regularity in the line-of-sight velocity of the observed emission features may first imply a periodic disc instability 
for producing these velocity-aligned molecular columns. 

Large-scale disc instabilities such as sausage and kink instabilities\cite{Griv2011} may not operate in a thin rotating disc 
surrounding an AGN but a gravitational instability can operate in a disc if its mass is comparable 
with the mass of the AGN itself\cite{KratterL2016}.
A gravitational instability may result in the formation of spiral arms in the disc and further gravitational contraction along these 
spiral arms may form distinct beads or clouds, as seen in the simulations of the Milky Way disc\cite{RenaudEA2013}.
The formation of dense spiral arms has indeed been put forward as an explanation of the high velocity features of NGC\,4258\cite{MaozM1998}.
When self-gravitation in the disc becomes important, the disc will become locally unstable for all axisymmetric perturbations 
when the Toomre stability coefficient $Q = (c_s \kappa)/(\pi G \Sigma) < 1$, 
where $c_s$ is the local sound speed, $\kappa = V / R_{em}$ is the epicyclic frequency (equal to $\Omega$ in a Keplerian disc), and
$\Sigma$ is the molecular surface density in the disc at the location of the emission regions\cite{Toomre1964,Toomre1981,KratterL2016}.
A simple estimate of the mass of a disc with an outer radius of 0.29 pc, with a uniform density of 
$n_d$ = $5 \times 10^{9} cm^{-3}$, and a scale height of 62 AU\cite{HerrnsteinEA2005,ArgonEA2007} gives a disc
mass $M_d$ = 1.9 $\times$ 10$^4$ \msol (Methods). 
This suggests a value for $Q$ = 24 for NGC\,4258 and indicates the disc to be very stable for gravitational instability.
Similar values of the disc mass have been derived from modelling the required disc accretion rate for driving the radio 
activity of the source\cite{MaozM1998,HerrnsteinEA2005,ModjazEA2005}.
Therefore, unless the disc is substantially more massive, a gravitational instability will not operate in NGC\,4258 and 
would not result in spiral arms with beads-on-a-string emission regions at the systemic velocity and in the tangential sections of the disc.

\begin{figure}[h]
\includegraphics[scale=0.38,angle=-90]{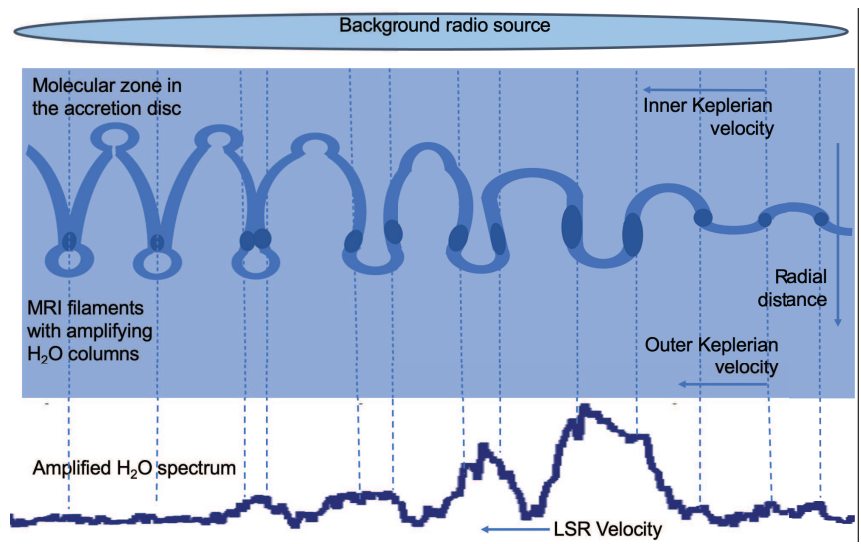}
\caption{{\bf Schematic cartoon of the non-linear development of a magneto-rotational instability and how velocity-coherent 
radial column densities may develop. }
The image attempts to reproduce the high-resolution RAGS18H data on the basis of the simulation results of the non-linear 
development of the MRI instability\cite{BalbusH1991}. 
The MRI waveform and the masering regions are moving with time from West to East in front of the 
nuclear continuum source, i.e. they are moving from low to high velocity or from right to left in the image.
The two radial sections in the flanks of the outward moving section of the MRI waveform may together provide the velocity-coherent 
path for background amplification for the high brightness features. 
The dark blue areas indicate the regions with an amplifying velocity-coherent radial column density and their relative size may explain the rapid rise of the features followed by a slower decrease with time.
The inward moving sections of the instability may contribute to the more diffuse emission seen in the total power spectra.
The size of the emission region is approximately half of the separation between the regions.
The formation of isolated MRI cell structures (towards the left) may suggest that features can remain in the total power spectra 
for a period of time.
\label{fig5}}
\end{figure}

Alternatively, the small-scale shear-driven interchange instability may be considered to facilitate the masering activity.
Such an interchange instability has been suggested early by Shakura and Sunyaev\cite{ShakuraS1973} to be the most likely 
viscosity agent in the thin Keplerian disc. 
Evidenced by the north-south extended radio structure and the non-thermal nuclear emission of NGC\,4258, 
there must be an active viscosity agent in the disc\cite{ArgonEA2007,DoiEA2013} causing an accretion rate on the 
order of $\dot M$ = 10$^{-4} \alpha$ \msols yr$^{-1}$ with a viscosity parameter $\alpha \leq 1$, 
which is determined on the basis of the radio and bolometric luminosity\cite{HerrnsteinEA2005}, 
the equipartition upper limit of the magnetic field\cite{ModjazEA2005}, and  the X-ray absorption within the 
disc\cite{FruscioneEA2005}.
The magneto-rotational version of this shear-drive interchange instability (MRI) can operate in a differentially rotating disc 
and would generate local 'sinusoidal' disruptions of the shear layers that interchange inner fluid elements with outer elements 
and cause viscosity and an exchange of momentum.  
This MRI operates independently of the strength of the B-field and varies linearly with the local value of 
the angular rotation velocity in the disc, where the field energy of a (self-amplified) poloidal (B$_z$) component does not
exceed the thermal energy density\cite{BalbusH1991,HawleyB1991} (see below). 
The presence of a weak poloidal B-field component serves to counteract Coriolis forces that would prevent non-linear growth 
in the non-magnetised version of this instability.
Large scale numerical simulations of the linear and non-linear evolution of the MRI show that the 'sinusoidal' toroidal 
field disruptions will develop into a series of radial interpenetrating 'fingers' of high and low angular momentum, that 
eventually break up into turbulence and strongly enhance the angular momentum transport\cite{HawleyB1991,BalbusHS1996}. 
These simulations show that MRIs would be able to operate in the disc of NGC\,4258 and that the scale size of the cells will be
on the order of the local scale height in the disc\cite{HoggR2017},

Phenomenologically the non-linear development of periodic MRI structures within the molecular zone of the disc provides 
an attractive scenario of creating radial column densities sufficient for H$_2$O masering action (Methods).
Simulations of the MRI waveform \cite{BalbusH1991,HawleyB1991} show that non-linear stretching of the radial segments 
of the initially sinusoidal  waveform could temporarily form a velocity-coherent molecular column density.
Amplification in these radial filaments with an enlarged optical depth could, temporarily, result in velocity-separated 
high-brightness emission features within the MRI waveform as depicted in the cartoon representation of Figure \ref{fig5}.
In this scenario for NGC\,4258, the two radial flanks of the outward moving MRI loop in the waveform form together 
an emission feature with a size of about half the separation between the regions and equal to the scale height in the disc, 
as observed.
Similarly, the contributions of two distinct flanks in the waveform would account for the phase changes observed in the 
middle of the profiles 
of some high resolution features of RAKS07AT and RAGS18H  (Figs. \ref{fig1} and \ref{fig3}).
However, following the initial formation of radial filaments in the MRI waveform, this waveform will further deform and disrupt 
the filaments and leave behind some turbulent cells, as evidenced by the MRI simulations\cite{BalbusH1991,HawleyB1991} and 
magnetohydrodynamic simulations of accretion discs\cite{HawleyGB1995}.
Considering that the observational data reveals only the presence of amplifying column densities within the MRI waveform, 
and not the waveform itself, the appearance in NGC\,4258 of a series of regularly spaced weak (remnant) features, preceded by 
transient high-brightness emission features provides a scenario that is consistent with the characteristic MRI behaviour.
Superpositions of multiple MRI strings may explain the observed spectra obtained during earlier observation epochs in NGC\,4258.
While the development of MRIs fits well with the current understanding of the accretion viscosity in the disc,
further study and simulations are needed to confirm the physical conditions required for the observed MRI development, and 
the formation of velocity-coherent radial molecular filaments.

The MRI scenario provides a natural explanation for the apparent periodicity of the spectral features on the basis of a viscous process 
that already may operate in the disc, and the sudden onset of non-linear development may explain the formation of radial column 
density enhancements and the observed flux evolution. 
The presence of two amplifying flanks in the MRI waveform may account for the variation of the observed line profiles and 
would be consistent with the phase changes in the middle of some high resolution features. 
Alternative scenarios for generating these same emission characteristics have been considered based on modulation 
of the interferometric fringe visibility amplitudes with projected baseline lengths, 
or bi-refringence, or special scattering/propagation conditions.
However, such scenarios also require a mechanism for generating sufficient (radial) path lengths of excited molecules in a 
differentially rotating and stratified medium, while a second mechanism would be needed for generating the periodic and flaring 
spectral behaviour. 
Although alternative scenarios other than beads-on-a-string spiral cloud structures or instabilities for generating the 
required column densities cannot be ruled out, no satisfactory alternative scenario has yet been devised to reproduce the 
observed emission characteristics.

Adopting a scenario with MRI-generated filamentary foreground structures amplifying a background radio continuum 
of the southern jet, more of the physical conditions of the emission regions can be estimated.
Because the radio continuum background at the base of the southern jet must be rather uniform with a flux 
density on the order of $S_c$ = 0.1 mJy at 22 GHz\cite{HerrnsteinEA1998,ArgonEA2007,DoiEA2013} (Methods), 
the amplifying optical depth associated with the flux density $S_{\ell}$ of the strongest cross-correlated feature in the RAGS18H 
spectrum can be estimated as follows from  $\tau_c = ln (S_{\ell}/ S_c) = -9.6$  (see Table \ref{tab1} in Methods). 
Slightly lower optical depth values are found for the other high resolution features in RAGS18H, while the 
values for the less resolved features in RAKS07AT and RAGS11AF become slightly higher.
The optical depth required for the peak in the auto-correlation spectrum in the RAGS18H data is $\tau_t$ = --11.3, which 
suggests that an additional halo contribution $\tau_d$ = --1.7 comes from the more diffuse toroidal sections of the MRI waveform. 
This diffuse contribution to the optical depth convolves the high brightness components in the auto-correlation spectra and 
decreases as the beam becomes larger at lower resolution.
Considering that the local density may be on the order of n(${H_2}$) = 10$^9$ cm$^{-3}$ with a water 
abundance n(H$_2$O)/n($H_2$) $\approx$ 10$^{-4}$ in the disc at 0.15 pc\cite{NeufeldMC1994,NeufeldM1995}, the 
actual population inversion in a (representative) 10 AU radial filament within an 84 AU sized emission region appears not 
very demanding at $\Delta n / n(H_2O) = 1.3 \times 10^{-7}$ (Methods).
The estimated brightness temperatures of the strongest masering components increase with a narrowing SVLBI beam 
from $T_b$ =  ~5.3 $\times$ 10$^{12}$ K for RAGS11AF, to 1.2 $\times$ 10$^{14}$ K for RAKS07AT, and to 2.5 $\times$ 10$^{14}$ 
K for RAGS18H.

Although magnetic fields are present in the disc of NGC\,4258, the strength of the field is not yet known.
While the MRI scenario does not depend directly on the magnetic field, a dominant toroidal component would be modified 
during the non-linear MRI development and produce a radial field component in the radial sections of the MRI waveform
(see Fig. \ref{fig5}). 
Similarly, the toroidal field component should be dominant for the high- and low-velocity features from the edge-on sections of the disc.
Magnetic fields have not yet been detected from within the disc and only an upper limit of 0.130 Gauss has been found for any 
radial field components associated with prominent features of NGC\,4258\cite{ModjazEA2005}.
Incidentally, a similar upper limit is found for the poloidal field component because the MRI only works when the energy density 
of the poloidal field component B$_z$ does not dominate the local thermal energy density. This suggests a ratio 
$\beta(B_z) = 8\pi\rho c^2/B_z^2 \geq 3$ \cite{BalbusH1991} and a value for B$_z$ $<$ 0.14 Gauss. 
Further evaluation is not yet possible for the effect of the magnetic fields on the conditions for MRI development and for 
the formation of the masering features.

{\bf 3. Summary}

In conclusion, the SVLBI observations of NGC\,4258 with the RadioAstron Observatory 
have resulted in new details about the H$_2$O masering action at the systemic velocity and the workings inside the rotating disc.  
At high spatial resolution, the single broad features observed in the total power spectra decompose into regularly spaced
high-brightness masering regions inside the disc.
These high-brightness regions detected with SVLBI are found to be part of a longer series of drifting and periodic emission 
regions that are also detected in total power data.
Interpreting the spectra obtained from these three high resolution data sets of NGC\,4258 shows that after an initial rapid 
growth in strength, the high-brightness features show a slow decrease in strength and eventually become 
part of this slowly drifting series of weak spectral features that remain in the spectrum for years. 

In terms of a maser amplification scenario, the observed emission features in NGC\,4258 only indicate the presence of 
regions with a radially amplifying column density of H$_2$O molecules where they do not directly explain the underlying 
mechanism producing these column densities.
Although more complex interpretations may be possible to interpret the apparently periodic and moving emission regions,  
their association with transiently varying radial sections in a magneto-rotational instability waveform provides an attractive scenario.
The sudden onset of the non-linear MRI development naturally explains the periodicity of the emission regions and the generation 
of a temporary enhancement of the molecular column density in the cells. 
The velocity-coherent column density containing excited H$_2$O molecules results in amplification of the background 
continuum, and further non-linear MRI development will in time deform the waveform and diminish/destroy this radial column density.
An MRI scenario for the emission regions suggests: 
first, that the size of the high-brightness features is roughly half the separation distance between the regions and equals the local scale height of the disc; 
second, that the observed velocity drifts confirm they are part of the differential rotation pattern in the disc; 
third, that the non-linear behaviour of the MRI explains the flaring and flux variability; 
and fourth, that the transient behaviour of the radial flanking sections of the MRI waveform may produce the required velocity-coherent amplifying optical depth. 
Furthermore, the association of MRI processes with the viscosity and turbulence causing radial momentum
 transfer provides a consistent link between the intensity of observed maser emission and the accretion flow in the differentially 
 rotating thin Keplerian disc.
In addition, this association  suggests that the accretion rate has been low during the current three epochs as compared with earlier 
epochs when a higher accretion rate with multiple MRI series filled the spectra with strong features.
The occurrence of an MRI in a Keplerian disc would indeed confirm this shear-driven instability to be an agent for generating the viscosity in the disc as proposed nearly 50 years ago by Shakura and Sunyaev.

The observed emission regions inside the molecular zone in the disc move from west to east across the line with 
the systemic velocity and within the MRI scenario the variation of their emission strength reflects the time-evolution of the 
velocity-coherent molecular columns in the radial filaments. 
The available observational data from earlier epochs needs to be re-investigated in order to discover patterns 
and variations of line fluxes that would further verify the currently emission scenario for the regions in the disc. 
Adopting an MRI scenario also for the observed redshifted and blueshifted edge-on disc features of NGC\,4258 would require that 
their amplifying optical depth follows from a tangential integration over a series of MRI cells. 
The velocity spacing of such features would then indicate the radial distribution of developed MRI structures in the disc
as it determines the ongoing viscosity process in the disc.

\begin{methods}

{\bf The RadioAstron observations -} The space VLBI observations of NGC\,4258 (M\,106) with the 10-metre 
radio telescope on the {\it RadioAstron Observatory}\cite{KardashevEA2013} in the time period between 
February 2014 and February 2017 have resulted in successful RAO experiments on baselines with 
ground-based radio telescopes (GRT).
The highest angular resolution on single baselines from RadioAstron to ground-based radio telescopes during experiments 
ranged  from 165 $\mu$as  for a perigee
space-Earth baseline of 1.0 ED (Earth diameter) to 7 $\mu$as fringe spacing for an apogee baseline of 26.9 ED. 
The three experiments presented in this paper demonstrate the effect of increasing 
resolution on determining the structure of the emission regions in NGC\,4258.
The observations were executed during relatively short time intervals of 1 to 1.5 hours in order to facilitate the necessary 
cooling periods for the space antenna, thus avoiding its deformation due to differential solar heating.
The ground-based facilities used for the H$_2$O Megamaser experiments with RadioAstron are Effelsberg (Germany), 
Green Bank (United States), Torun (Poland), Yebes (Spain), and the Kwazar stations Kalyazin, Svetloe, and Badary (Russia).

The conversion from angular resolution to spatial resolution is 7.62 AU/micro-arcsecond at a distance of 7.6 Megaparsecs.

The RadioAstron experiments described here employed the 22 GHz receiver system operating within a fixed
frequency window of 22188 - 22204 MHz, with an estimated T(system) = 100 - 127 K.
The observation of the H$_2$O vapour "6(1,6)-5(2,3) F=5-4"  transition used a rest frequency of 22 235.120 MHz. 
The location of the redshifted spectral emission of NGC\,4258 at 22 199.96 MHz within the 16 MHz frequency 
window varies with the relative velocity of the spacecraft during its orbit at the 
time of the observations, and is often close to the edge of the spectral window.

The diameter of the RadioAstron antenna of only 10 m is small compared with the large 100 m class ground-based 
facilities such as the Green Bank Telescope (GBT) and the Effelsberg Telescope (EFF).
However, since the sensitivity on a VLBI baseline depends on the product of the two antennas, the VLBI 
measurements with RadioAstron are also most sensitive on the longest baselines when using one of these large ground-based telescopes.
RadioAstron baselines with smaller ground-based antennas are less sensitive but they are essential for verification and calibration purposes.
Because the RAO baselines are much longer than any of the terrestrial baselines, the results reported on in this paper are 
single baseline measurements and cannot be used for mapping the source.

Experiment RAKS07AT on a baseline of 9.1 - 9.8 ED was executed for 60 minutes on 5 February 2014 (2014.0986) 
with ground stations at Effelsberg (EFF, Germany; RAO-EFF = 9.4 ED), Torun (TR, Poland; RAO-TR = 9.1 ED), and Badary 
(BD, Russia; RAO-BD = 9.8 ED), resulting in space-ground and ground-ground fringes having an angular resolution of 
23 $\mu$as on an Earth-space baseline at 22 GHz, which corresponds to 175 AU at NGC\,4258.
A bandpass calibration was performed using a 5 minute observation of the calibrator source 1219+044.

Experiment RAGS11AF on baselines 0.5 - 1.9 ED was successfully executed for 60 minutes on 18 December 2014 
(2014.964) with ground stations at Green Bank (GBT, USA; RAO-GBT = 1.9 ED) and Torun (TR, Poland; RAO-TR = 1.4 ED).
Space-ground fringes resulted in an angular resolution of 113 $\mu$as on an Earth-space baseline at 22 GHz 
corresponding to 861 AU at the target source.
It was not possible to use the 5 minute observation of the calibrator source 1150+497 intended for bandpass 
calibration of the target data. 
Instead, an offline baseline fitting was used to remove some of the baseline structure in Figure \ref{fig2}.

Experiment RAGS18H on a baseline of 19.5 ED was executed for 74 minutes on 17 March 2016 (2016.2103) with 
ground stations at Green Bank (GBT, USA), Yebes (YB, Spain), and Torun (Poland). 
Space-ground fringes provide an angular resolution of 11 $\mu$as at 22 GHz, corresponding to 84 AU for NGC\,4258. 
Ground-ground fringes were obtained on the terrestrial GBT - TR baseline.
A 15-minute observation of the continuum calibrator source 0851+202 has been used for bandpass calibration.

{\bf Data processing -- } 
The staff at the ASCFX correlator at the Astrospace Center in Moscow has correlated the spectral line data 
observations multiple times in order to find and confirm interferometric fringes on the Earth-space 
baselines\cite{LikhachevEA2017}. 
The spectra of the full 16 MHz correlated bandwidth are presented here.
The data for bandpass calibrator sources were correlated using the same channel spacing as the spectral line data 
for RAKS07AT and RAGS18H. For RAGS11AF no bandpass calibrator was available and a baseline fitting routine was 
applied for the auto-correlation spectrum of the Effelsberg data.
Finding interferometric fringes on the baselines with the orbiting antenna is challenging because of
 the typically low signal-to-noise ratio (SNR) of the fringes and the positional uncertainties in the spacecraft orbit.
During the post-correlation, a cursory fringe search for the delay and delay rate (to find the distance and velocity 
of the spacecraft) was performed with the PIMA software\cite{PetrovEA2011} for processing individual baselines. 
The PIMA searches resulted in fringe detections up to baseline lengths of 26.9 ED for NGC\,4258. 
After setting the instrumental delays of the system, a second step of fringe-fitting was used to 
update the delay and delay rate model using the task {\it FRING} from the Astronomical Image Processing System (AIPS). 
Although this task does allow combining data from multiple ground telescopes in a global solution, it also serves well for 
our single baseline detection experiments where no imaging can be possible. 
Further data reduction has been done with AIPS using the tasks {\it BPASS} for bandpass calibration with data of 
nearby calibrators, and {\it APCAL} for amplitude calibration using the {\it ANTAB} file with station gain curves 
and system temperatures. 
The velocity scale of the correlated signals was determined from the actual channel frequencies and the known 
frequency offsets during the observations. 
In addition, the radial component of the observer's velocity (i.e. the ground station) at the time of observation was 
used to adjust the velocity scale of the spectra as described for the AIPS task {\it CVEL}.
All auto-correlation and single baseline cross-correlation Stokes $I$ spectra and phases were obtained 
with the task {\it POSSM}, and exported for plotting and smoothing in a {\it MATLAB} package. 

{\bf Structure of the background radio continuum --} 
Within a maser amplification scenario both the foreground optical depth and the background radio continuum are 
the important parameters.  
Therefore, strong emission from a moving region in the accretion disc may also result from a local enhancement 
in the radio continuum background.
In the disc configuration of NGC\,4258, a location for the radio continuum enhancement west of the transit position in the disc will 
result in strong emission features below the systemic velocity. 
If the location is east, all strong features would be above the systemic velocity.
However, for NGC\,4258 the strong features are found both below and above the systemic velocity, which suggests 
a rather uniform radio background without substantial flux enhancements.

{\bf The Velocities of the Masering features and the Drift Rate --} 
The characteristics of the features have been determined by measuring the spectrum and by Gaussian fitting procedures of the convolved features. 
The main objective for these measurements was to determine the centroid velocities, while the amplitudes and line-widths of the 
resolved features are not yet important for this evaluation.
Gaussian fitting of isolated features may not be adequate because of baseline structures and non-Gaussian shapes of the lines. 
Adequate Gaussian fitting of the convolved main profiles in the total power data and also the convolved profile in the cross-correlated 
RAGS11AF profile would require a priori knowledge about the velocities of the underlying components. 
In the case of RAKS07AT and RAGS18H, the number of components and their centre velocities may be derived from the cross-correlated 
spectra, which may be further verified from the breaks in the phase information. 
Only for the partially resolved cross-correlated spectrum as well as the auto-correlated spectra of RAGS11AF, three underlying 
components were assumed to be aligned with the regular pattern in the other parts of the spectrum and guided by breaks in the
phases of the cross-correlated data.
Considering the uncertainties in determining the velocities of the features, they appear regularly spaced although 
some irregularity may be seen in the separation between features at higher velocities.
The component velocities and their estimated strength are presented in Table \ref{tab1}.

All identified velocity components in the spectral data at the three epochs are displayed in the diagram in Figure \ref{fig4}, 
and the high brightness components seen in the cross-correlation data have been indicated.
The time intervals between the data of RAKS07AT, RAGS11AF and RAGS18H are 0.865 years and 1.244 years.
Considering that there are substantial time intervals between these observational epochs, linear extrapolations 
have been indicated to show that the pattern of weak and strong velocity components repeats from epoch to epoch.
Inaccuracies in the velocity determinations resulting from changes in the emission profiles and total power baselines structures 
add uncertainty to the drift rate determined from these data.
Since the emission features are associated with a non-linearly developing instability, some of these data points may show a velocity offset and 
some may have disappeared in time.
A mean value of the drift rate starting at RAKS07AT and ending at RAGS18H is found to be 11.1$\pm$0.5 \kms yr$^{-1}$, which is indeed close to 
the upper end of the range of 6.2 to 11.6 \kms yr$^{-1}$ observed during earlier monitoring campaigns\cite{HaschickB1994,HumphreysEA2008}.
Choosing different sequences of three data points (up or down) for linear extrapolations at subsequent epochs results in 
impossible fits and unrealistic values for the drift rate.
Tracing the features across the epochs also confirms the intensity evolution of individual velocity features during the 2.1-year time interval.
After a long quiescent period of the MRI with no emission, the maser features show a rapid increase in strength followed by a slower decay, 
after which they remain for years as remnant features in the spectrum.
This evolution is shown by the decay of the strong features of RAKS07AT and RAGS11AF and by the lowest velocity features 
in the RAGS11AF evolving into the first three prominent features in the shifted spectral window of the RAGS18H observation 
(Figs. \ref{fig1}-\ref{fig3}). 
This evolution has also been depicted in the cartoon representation of the non-linear MRI development in Figure \ref{fig5}.

{\bf The mass of the disc and the accretion rate --}
When self-gravitation in the disc becomes important, the disc will be locally unstable and starts forming spiral arms 
when the Toomre stability coefficient:
\begin{equation} 
Q = (c_s \Omega_k)/(\pi G \Sigma)  
\end{equation}
\noindent becomes less than unity\cite{Toomre1964,KratterL2016},
where $c_s$ is the local sound speed, $\Omega_k = V / R_{em} = \sqrt (G M_{bh} / R_{em}^3)$ at the emission regions,
the molecular surface density in the disc $\Sigma = 2 H \times n(H_2) \times 2 m_{p}$, and $R_{em}$ is the radius and $H$ is 
the scale height at the location of the emission regions.
For a local temperature of 800K the speed of sound is estimated at $c_s$ = 3.3 \kmss, 
the local density $n(H_2)$ = $5 \times 10^{9} cm^{-3}$, 
and the orbital velocity in the Keplerian disc is 1171 \kms at a radius $R_{em}$ = 0.126 pc.
This gives an estimate of the value for the Toomre stability parameter $Q$ = 24, which suggests the disc is rather stable 
against gravitational instabilities.
A simple estimate of the mass of an accretion disc with outer radius of 0.29 pc with a uniform density of 
$n_d$ = $5 \times 10^9 cm^{-3}$ and a scale height of 62 AU\cite{HerrnsteinEA2005,ArgonEA2007} gives a disc
mass $M_d$ = 1.9 $\times10^4$ \msols.
Similar values may also be derived from models of the required disc accretion rate on the order of 
$\dot M$ = $10^{-4} \alpha$ \msol{} $yr^{-1}$  ($\alpha \leq 1$) 
to drive the radio activity of the source\cite{MaozM1998,HerrnsteinEA2005,ModjazEA2005}.
Typically for a mass ratio $M_{d}/M_{bh} \leq 10^{-2}$, the self-gravity effect is not important\cite{KratterL2016,KuoEA2018} 
and this ratio would be $5\times10^{-4}$  for NGC\,4258.
These stability estimates suggest that the accretion disc in NGC\,4258 is gravitationally stable and that the formation of spiral 
structures, spiral shock waves and beads-on-a-string is not very likely.

{\bf Behaviour of a Magneto-Rotational Instability --}
The association of apparently periodic and moving emission regions with transiently varying radial sections in a 
magneto-rotational instability (MRI) waveform provides an attractive scenario.
Starting with the idea that MRI operates in accretion discs and is the cause of the necessary viscosity in the disc, 
this instability naturally explains the periodicity for the emission regions, and can create a temporary molecular column density in 
the cells, and, as simulations show, there can be a sudden onset of non-linear development. 
A velocity-coherent column density containing excited water molecules will result in temporary amplification of the background 
continuum, but further non-linear MRI development will in time deform the waveform and diminish/destroy this radial column density.

The MRI interchange instability only operates in the presence of a weak poloidal field\cite{BalbusH1991} that is needed to 
compensate for the Coriolis forces that would otherwise prevent non-linear growth in the non-magnetised version of the instability 
as proposed early by Shakura and Sunyaev\cite{ShakuraS1973}. 
Although the MRI will operate independently of the strength of the local magnetic field, there is a self-regulated limit for the 
poloidal B-field component to be less than the local energy density.
Although the MRI may be affected by the magnetic field conditions, yet it does not put any requirements on the radial and 
toroidal B-field components.

The MRI simulations presented in the series of papers by Balbus and Hawley\cite{BalbusH1991,HawleyB1991,BalbusHS1996}
show that MRI waveforms can develop in a disc over multiple orbits and then suddenly show a non-linear increase in amplitude.
During this sudden non-linear development the sinusoidal cells will be stretched radially and form radial strings of swept material.
The MRI structure could be visualised as a 2D structure in radial and toroidal directions in such a way that the vertical (poloidal) 
height of the cell is equal to the scale height of the disc\cite{HoggR2017}. 
As a result the dominant molecular columns during a non-linear development will extend perpendicular to the disc 
and they are equally spaced along the centreline in the toroidal (and velocity) direction. 
The differentially rotating atmospheric surface layer of Jupiter shows similar shear-driven structures. 
An amplifying column density only works when the excited molecules in the column are velocity-coherent, which suggests that 
this velocity coherence gets destroyed easily during further non-linear evolution of the MRI.
The simulations also show that the end result of the non-linear MRI development phase will produce blobs of entrained gas 
alternatingly moving inward and outward, which will account for radial momentum transfer in the disc and the generation of viscosity
in the disc.
 Interpreting the results of the simulations\cite{BalbusH1991}, a cartoon has been made to visualise the evolution of the 
 instability and the change of the amplifying column density in Figure \ref{fig5}. 
 In this scenario, the amplifying gains along each of the filaments will determine the shape of the feature, such that high optical depths 
 produce a single peaked feature and smaller optical depths can produce flat-shaped (maybe double-peaked) features.
  
{\bf Masering Conditions in the Disc --}
The optical depths deduced for the emission lines detected in our experiments appear very large because the radio continuum 
background associated with the base of the southern jet is very low in NGC\,4258.
However, the expected path length with a coherent velocity can be very large for an MRI structure in a parsec scale accretion disc.
The optical depth or maser gain for the H$_2$O column density may be expressed as\cite{Elitzur1992}:
\begin{equation}
\tau = (h \nu/ 4 \pi \Delta v_D)  g_2 B_{21} \int (n_2 - n_1) d\ell,
\end{equation}
\noindent where $\nu$ is the transition frequency, $\Delta v_D$ is the Doppler line width, $g_2$ is the statistical weight of the 
upper sub-level and $B_{21}$ is the Einstein absorption coefficient.
The difference in the upper and lower energy level population is expressed as $\Delta n = n_2 - n_1$.
For the 22 GHz $6_{16} - 5_{23}$ transition of water vapour, this expression reduces to 
$\tau = -5.02 \times 10^{-12} (\Delta n \ell)$ with $\ell$ being the path length.
For a negative optical depth of $\tau$ = -10, the required column density of inverted molecules 
$(\Delta n \ell) = 2 \times 10^{12} cm^{-1}$
For a particle density of $N_{H_2} = 10^9 cm^{-3}$ and a relative H$_2$O abundance of 10$^{-4}$ in the disc 
at 0.15 pc\cite{NeufeldMC1994,NeufeldM1995}, the required inversion level 
$(\Delta n /n_{H_2O}) = 1.3 \times 10^{-6} AU^{-1}$. 
For a possible path length of 10 AU, the inversion level along a velocity coherent 
path needs to be $1.3 \times 10^{-7}$, which would be very low for the environment in the molecular zone. 

\end{methods}

\begin{addendum}

\item{}  
 The authors dedicate this paper to the memory of our colleague and friend Nikolai Kardashev, a man of great vision, who 
 persevered to realise the RadioAstron mission.
  
The authors thank the observatory staff of the ground telescope stations Effelsberg, Green Bank, Torun, Yebes, 
and the Kwazar stations Kalyazin, Svetloe, and Badary for their participation in the observations.
These observations have been correlated at the ASC DiFX correlator  and the authors thank the 
Correlator Team members for their contributions, their repeated re-correlation efforts, and their unfailing support of this project.
The authors also thank the other members of the H$_2$O Megamaser Team for their support for this project:
Alexei Alakoz, Simon Ellingsen, Ivan Litovchenko, James Moran, and Alexander Tolmachev.
The authors thank Eduard Vorobyov (Uni. Vienna) for valuable discussions about the stability criteria for the disc.

The RadioAstron project has been led by the AstroSpace Centre of the Lebedev Physical Institute of the 
Russian Academy of Sciences and the Lavochkin Scientific and Production Association under a contract 
with the State Space Corporation ROSCOSMOS, in collaboration with partner organisations in Russia and other 
countries. 
The National Radio Astronomy Observatory is a facility of the National Science Foundation operated under
a cooperative agreement by Associated Universities, Inc. 
The European VLBI Network is a joint facility of independent European, African, Asian, and North American 
radio astronomy institutes. 

WAB acknowledges the support from the National Natural Science Foundation of China under grant No.11433008 
and the Chinese Academy of Sciences President's International Fellowship Initiative under grants No. 2019VMA0040,
2021VMA0008 and 2022VMA0019. 
TA acknowledges the grant support from the Youth Innovation Promotion Association of CAS. 
AMS was supported by the Ministry of Education and Science of Russia (the basic part of the State assignment, 
K 1567 no. FEUZ-2020-0030)

\item[Author contributions ] 
 WAB coordinated the research, carried out the data reduction, and wrote the manuscript.
WAB, TA, CH, HI, VK, and AS contributed to the discussion and interpretation of the data, and provided comments on the manuscript.
WAB served as Principal Investigator of the {\it RadioAstron} MegaMaser Key Science Program.

\item[Competing Interests] 
The authors declare that they have no competing financial interests.

\item[Data availability and Code availability ]
 The correlated data for RadioAstron experiments RAGS11AF, RAKS07AT and RAGS18H are available from 
the RadioAstron Data Archive at the Astrospace Center of the PN Lebedev Physics Institute ASC LPI) in Moscow at 
http://opendata.asc.rssi.ru/index.php.

Data reduction has been done with the Astronomical Image Processing System, AIPS, that has been developed by 
the National Radio Astronomy Observatory and is documented and available at 
\newline http://www.aips.nrao.edu/index.shtml.

Plotting procedures have been used from the MATLAB Toolbox distributed by MathWorks, Inc. at 
\newline https://www.mathworks.com/products/matlab.html.

The PIMA software package for processing individual baselines can be found at http://astrogeo.org/pima/. 

A description of the ASC Software correlator can be found in the literature\cite{LikhachevEA2017}.

\item[Correspondence] 
 Correspondence should be addressed to WAB (baan@astron.nl).

\end{addendum}

\begin{table*}[h!]
\footnotesize
\centering
  \caption{Spectral Components of NGC\,4258}
  \begin{threeparttable}
   \begin{tabular}{ccccc}
    \hline\hline
     Name  & Baseline & Component & Total     & Cross \\
      Epoch     &           &                     &  Power  & Power   \\
         (yr)       & (ED)   &   (km/s)\tnote{1}  &   (Jy)\tnote{1}     & (Jy)\tnote{1}        \\
                   \hline                                      
RAKS07AT            & 9.5 & 470.3$\pm$0.3 & 2.31$\pm$0.03 & 0.37$\pm$0.24 \\
2014.0986                    &  & 476.8 & 11.28 & 4.28 \\
EFF -RAO			   &  & 483.3 & 7.17 & 2.51 \\		   
				    &  & 489.8 & 3.07 & 0.65 \\
				    &  & 499.0 & 0.52 & --  \\          
				    &  & 507.0 & 0.56 & -- \\
				    &  & 515.0 & 0.46 & -- \\
				    &  & 523.0 & 0.38 & -- \\		
				    \hline	    
  RAGS11AF               & 1.3 & 460.0$\pm$0.3& 0.41$\pm$0.04& 0.13$\pm$0.07\\
  2014.964		& &  472.3 & 0.39 & 0.23  \\
  GBT-RAO 	         &  &  478.8 & 1.41 & 1.01  \\
  	                         &  & 485.2 & 9.88  & 5.55 \\
                                  &    &  493.3 & 0.95 &  0.66 \\                      
                                    &   &  501.4 & 0.26 &  0.22\\               
                                         &  &  510.1 & 0.24 & 0.10 \\
                                      &  & 517.5  & 0.24 & 0.13  \\ 
 				   &  &  526.5 & 0.46 & 0.18  \\
                                     &  &  533.3 & 0.45 & 0.09  \\
                                      &  &  540.5 & 0.41 & 0.13  \\
\hline
RAGS18H          	&  19.5 & 450.4$\pm$0.3 & 1.40$\pm$0.02 & 0.21$\pm$0.11\\                              
 2016.2082	        &  & 457.0 &  1.80  & 0.20 \\
 GBT-RAO		&  & 464.2 & 8.20 &1.50\\
				&  & 471.5 & 5.60 & 1.12\\
				&  & 478.9& 2.40 & 0.33\\
				&  & 486.2 & 1.20 & 0.26\\
				&  &  493.5 & 0.36 & -- \\
				 &  & 500.3  & 0.21 &  -- \\
				 &  &  509.0 & 0.23 &  -- \\				                                         
				&  &  521.0 & 0.27 & -- \\		
				&  & 535.2  & 0.30 & -- \\
				&  &  560.0  & 0.07 & -- \\		                                        
	                   \hline
     
         \end{tabular}
         \begin{tablenotes}
         \item[1]{Note 1: The three sigma errors apply to all components for each of the observations.
         The errors in velocity reflect the estimated measurement errors and the amount of spectral smoothing applied.
         The error equals three times the 0.105 \kms channel width of the correlated data.
         The errors in amplitude reflect the three sigma standard deviation of the featureless sections of the spectra. }
   	 \end{tablenotes}
	\label{tab1}
	   \end{threeparttable}
	   	   \end{table*}

\newpage {\bf Figure Captions}

\end{document}